\documentclass[prl,superscriptaddress,twocolumn,showpacs]{revtex4}
\usepackage{epsfig}
\usepackage{graphicx}
\usepackage{amsmath}
\usepackage{bm}
\usepackage{amssymb}
\usepackage{color}

\definecolor{purple}{rgb}{0.8,0,0.6}
\definecolor{darkgreen}{rgb}{0.00,0.6,0.00}

\begin{document}


\title{Consistent Chiral Kinetic Theory in Weyl Materials: Chiral Magnetic Plasmons}
\date{\today}

\author{E.~V.~Gorbar}
\affiliation{Department of Physics, Taras Shevchenko National Kiev University, Kiev, 03680, Ukraine}
\affiliation{Bogolyubov Institute for Theoretical Physics, Kiev, 03680, Ukraine}

\author{V.~A.~Miransky}
\affiliation{Department of Applied Mathematics, Western University, London, Ontario N6A 5B7, Canada}
\affiliation{Department of Physics and Astronomy, Western University, London, Ontario N6A 3K7, Canada}

\author{I. A.~Shovkovy}
\affiliation{College of Integrative Sciences and Arts, Arizona State University, Mesa, Arizona 85212, USA}
\affiliation{Department of Physics, Arizona State University, Tempe, Arizona 85287, USA}

\author{P.~O.~Sukhachov}
\affiliation{Department of Applied Mathematics, Western University, London, Ontario N6A 5B7, Canada}

\begin{abstract}
We argue that the correct definition of the electric current in the chiral kinetic theory for Weyl materials
should include the Chern-Simons contribution that makes the theory consistent with the local conservation
of the electric charge in electromagnetic and strain-induced pseudoelectromagnetic fields. By making use
of such a kinetic theory, we study the plasma frequencies of collective modes in Weyl materials in constant
magnetic and pseudomagnetic fields, taking into account the effects of dynamical electromagnetism.
We show that the collective modes are chiral plasmons. While the plasma frequency of the longitudinal
collective mode coincides with the Langmuir one, this mode is unusual because it is characterized not only
by oscillations of the electric current density, but also by oscillations of the chiral current density. The latter
are triggered by a dynamical version of the chiral electric separation effect. We also find that the plasma
frequencies of the transverse modes split up in a magnetic field. This finding suggests an efficient means of
extracting the chiral shift parameter from the measurement of the plasma frequencies in Weyl materials.
\end{abstract}



\maketitle

{\em Introduction.---}
The study of the fundamental properties of magnetized relativistic matter attracted a lot of attention in
recent years. The physical systems in question include the plasmas in the early Universe \cite{Kronberg} and
relativistic heavy-ion collisions \cite{Kharzeev:2008-Nucl,Kharzeev:2016}, degenerate states of dense matter
in compact stars \cite{Kouveliotou:1999}, and a growing number of recently discovered three-dimensional
Dirac and Weyl materials \cite{Borisenko,Bian,Long}. To large extent, the recent increased activity in the
studies of magnetized relativistic matter is driven by the hope of detecting macroscopic implications of
quantum anomalies. One of such implications is the celebrated chiral magnetic effect (CME)
\cite{Kharzeev:2008}, which has been detected indirectly in the quark-gluon plasma created in
heavy-ion collisions (for a review, see Ref.~\cite{Kharzeev:2016}), as well as in Dirac semimetals
\cite{Li-Kharzeev:2016}. Note that the interpretation of the heavy-ion experiments is not without
a controversy \cite{CMS:2016}.

The search for macroscopic implications of quantum anomalies is greatly facilitated by the recent discovery of
Dirac and Weyl materials, whose low-energy quasiparticle excitations are described by relativistic-like equations.
Indeed, unlike most forms of truly relativistic matter, these novel condensed matter materials open the possibility for
revealing and testing many anomalous effects in magnetized matter in table-top experiments under controlled
conditions. Moreover, they may even allow for modeling phenomena that are impossible in relativistic physics.
A specific example is provided by a background pseudomagnetic (or, equivalently, axial magnetic) field
$\mathbf{B}_5$, which can be effectively produced by a mechanical strain in Dirac and Weyl materials
\cite{Zubkov:2015,Cortijo:2016yph,Grushin-Vishwanath:2016, Pikulin:2016}. In essence, the pseudomagnetic field
$\mathbf{B}_5$ resembles the ordinary magnetic field $\mathbf{B}$, but acts on opposite chirality
quasiparticles so as if they had opposite charges. In the case of the Dirac semimetal Cd$_3$As$_2$,
for example, the estimated strength of the strain-induced pseudomagnetic field could range from about $B_5\approx0.3~\mbox{T}$
in twisted nanowires \cite{Pikulin:2016} to $B_5\approx15~\mbox{T}$ in bended thin films \cite{Liu-Pikulin:2016}.
Similarly, a pseudoelectric field $\mathbf{E}_{5}$ can be generated by time-dependent deformations.

Collective excitations are simple, but informative probes of plasma properties \cite{Landau:t10}. It is natural to ask,
therefore, whether such modes in chiral plasmas can be affected by quantum anomalies. The authors of Ref.~\cite{Kharzeev}
proposed that the chiral anomaly implies the existence of a new type of collective excitation, i.e., the chiral magnetic wave (CMW),
that originates from an interplay of chiral and electric charge density waves. In this study, we will investigate the collective
modes in Weyl materials, using the framework of the chiral kinetic theory with a proper treatment of dynamical electromagnetism.

The central idea of this Letter is to use the correct definition of the electric current in the chiral kinetic
theory for Weyl materials with strain-induced pseudoelectromagnetic fields. As we show, the current should
necessarily include the Chern--Simons contribution, which is also known as the Bardeen-Zumino polynomial
\cite{Bardeen}. Such a correction restores the local conservation of the electric charge in the case of
general electromagnetic and pseudoelectromagnetic fields. In addition, this topological term affects the
properties of collective modes. For example, their plasma frequencies acquire a dependence on the chiral
shift parameter, i.e., the momentum-space separation between the Weyl nodes.

{\em Model.---} The chiral kinetic theory is a semiclassical theory, which describes the time evolution of the
one-particle distribution functions $f_{\lambda}$ for the right- ($\lambda=+$) and left-handed ($\lambda=-$) fermions.
In the collisionless limit (assuming that the frequency of collective excitations $\omega$ is much larger that
inverse relaxation time $1/\tau$), the kinetic equations are given by \cite{Stephanov,Son}
\begin{eqnarray}
&&\frac{\partial f_{\lambda}}{\partial t}+
\frac{\left[e\tilde{\mathbf{E}}_{\lambda}
+\frac{e}{c}(\mathbf{v}\times \mathbf{B}_{\lambda})
+\frac{e^2}{c}(\tilde{\mathbf{E}}_{\lambda}\cdot\mathbf{B}_{\lambda})\mathbf{\Omega}_{\lambda}\right]\cdot
{\bm{\nabla}_{\mathbf{p}} f_{\lambda}}}{1+\frac{e}{c}(\mathbf{B}_{\lambda}\cdot\mathbf{\Omega}_{\lambda})} \nonumber\\
&&+\frac{\left[\mathbf{v}+e(\tilde{\mathbf{E}}_{\lambda}\times\mathbf{\Omega}_{\lambda})
+\frac{e}{c}(\mathbf{v}\cdot\mathbf{\Omega}_{\lambda})\mathbf{B}_{\lambda}\right]
\cdot\bm{\nabla}_{\mathbf{r}} f_{\lambda}}{1+\frac{e}{c}(\mathbf{B}_{\lambda}\cdot\mathbf{\Omega}_{\lambda})}=0,
\label{CKT-kinetic-equation}
\end{eqnarray}
where $\mathbf{E}_{\lambda}=\mathbf{E}+\lambda\mathbf{E}_{5}$ and $\mathbf{B}_{\lambda}=\mathbf{B}+\lambda\mathbf{B}_{5}$
are effective electric and magnetic fields for fermions of chirality $\lambda$,
$\mathbf{\Omega}_{\lambda} =\lambda \hbar \mathbf{p}/(2p^3)$ is
the Berry curvature \cite{Berry:1984}, $p\equiv|\mathbf{p}|$,
$\tilde{\mathbf{E}}_{\lambda} = \mathbf{E}_{\lambda}-(1/e)\bm{\nabla}_{\mathbf{r}} \epsilon_{\mathbf{p}}$,
and the factor $1/[1+e(\mathbf{B}_{\lambda}\cdot\mathbf{\Omega}_{\lambda})/c]$ accounts
for the correct phase-space density of chiral states in an effective magnetic field  \cite{Xiao}. By making use of the
fermion dispersion relation, valid up to the linear order in the background magnetic field $\mathbf{B}_{\lambda}$ \cite{Son},
\begin{equation}
\epsilon_{\mathbf{p}}= v_Fp\left[1 - (e/c)(\mathbf{B}_{\lambda}\cdot \mathbf{\Omega}_{\lambda})\right],
\label{CKT-epsilon_p}
\end{equation}
we derive the quasiparticle velocity $\mathbf{v}=\bm{\nabla}_{\mathbf{p}} \epsilon_{\mathbf{p}}$, i.e.,
\begin{equation}
\mathbf{v}=v_F\frac{\mathbf{p}}{p} \left[1+2\frac{e}{c} \left(\mathbf{B}_{\lambda} \cdot \mathbf{\Omega}_{\lambda}\right) \right]
- \frac{e v_F}{c p}\mathbf{B}_{\lambda}\left(\mathbf{p} \cdot \mathbf{\Omega}_{\lambda}\right).
\label{CKT-v}
\end{equation}
Here $v_F$ is the Fermi velocity.

The equilibrium distribution functions for chiral fermions are given by the standard Fermi-Dirac
distributions
\begin{equation}
f^{\rm (eq)}_{\lambda}=\left[e^{(\epsilon_{\mathbf{p}}-\mu_{\lambda})/T}+1\right]^{-1} ,
\label{CKT-equilibrium-function}
\end{equation}
where $T$ is the temperature (measured in energy units) and $\mu_{\lambda}=\mu+\lambda\mu_5$
are the effective chemical potentials for the right- and left-handed fermions. Note that $\mu$ and $\mu_5$
are the electric and chiral chemical potentials, respectively. The distribution functions for antiparticles
$\bar{f}^{\rm (eq)}_{\lambda}$ are obtained by replacing $\mu_{\lambda}\to - \mu_{\lambda}$. In addition,
for antiparticles, one should replace $e\to-e$ and $\mathbf{\Omega}_{\lambda}\to-\mathbf{\Omega}_{\lambda}$.

The charge and current densities are given by \cite{Son}
\begin{eqnarray}
\label{CKT-charge}
\rho_{\lambda} &=& \sum_{\rm p,a}e\int\frac{d^3p}{(2\pi \hbar)^3}
\left[1+\frac{e}{c}(\mathbf{B}_{\lambda}\cdot\mathbf{\Omega}_{\lambda})\right]f_{\lambda},\\
\label{CKT-current}
\mathbf{j}_{\lambda} &=& \sum_{\rm p,a}e\int\frac{d^3p}{(2\pi \hbar)^3}\left[\mathbf{v}
+\frac{e}{c}(\mathbf{v}\cdot\mathbf{\Omega}_{\lambda}) \mathbf{B}_{\lambda}
+e(\tilde{\mathbf{E}}_{\lambda}\times\mathbf{\Omega}_{\lambda})\right]f_{\lambda} \nonumber\\
&&+\sum_{\rm p,a}e\bm{\nabla}\times \int\frac{d^3p}{(2\pi \hbar)^3} f_{\lambda}\epsilon_{\mathbf{p}}\mathbf{\Omega}_{\lambda},
\end{eqnarray}
where $\sum_{\rm p,a}$ denotes the summations over particles and antiparticles and the last term describes a magnetization current.

{\em Local charge nonconservation.---} By using Eqs.~(\ref{CKT-kinetic-equation}), (\ref{CKT-charge}), and (\ref{CKT-current}) together with the Maxwell's equations, one can easily
derive the following continuity equations for the chiral and electric currents:
\begin{eqnarray}
\label{CKT-dn/dt-n5}
\frac{\partial \rho_5}{\partial  t}+\bm{\nabla}\cdot\,\mathbf{j}_5 &=& \frac{e^3}{2\pi^2 \hbar^2 c}
\Big[(\mathbf{E}\cdot\mathbf{B}) +(\mathbf{E}_{5}\cdot\mathbf{B}_{5})\Big],\\
\label{CKT-dn/dt-n}
\frac{\partial \rho}{\partial  t}+\bm{\nabla}\cdot\,\mathbf{j} &=& \frac{e^3}{2\pi^2 \hbar^2 c}
\Big[(\mathbf{E}\cdot\mathbf{B}_{5}) +(\mathbf{E}_{5}\cdot\mathbf{B})\Big].
\end{eqnarray}
The first equation is related to the celebrated chiral anomaly \cite{ABJ} and expresses the nonconservation
of the chiral charge in the presence of electromagnetic or pseudoelectromagnetic fields. Physically,
this nonconservation can be understood as pumping of the chiral charge between the Weyl nodes of opposite
chiralities. The second equation describes the anomalous local nonconservation of the {\em electric} charge
in electromagnetic and pseudoelectromagnetic fields.

The local nonconservation of the electric charge is a very serious problem. If taken at face value,
it would imply that the electric charge is literarily created out of nothing. It was suggested in
Refs.~\cite{Pikulin:2016,Grushin-Vishwanath:2016} that it may correspond to pumping of the
charge between the bulk and the boundary of the system. However, it is unclear how such a
spatially nonlocal process could resolve the problem.

As we argue below, the resolution of the problem is much simpler. It lies in the fact  that
Eqs.~(\ref{CKT-dn/dt-n5}) and (\ref{CKT-dn/dt-n}) are the so-called {\it covariant} anomaly relations
that come from the fermionic sector of the theory, in which left- and right-handed
fermions are treated in a symmetric way. Just like in quantum field theory, this is inconsistent with
the gauge symmetry. The correct physical currents, satisfying the local conservation of the electric
charge, are the {\it consistent} currents \cite{Landsteiner:2013sja}. A very clear discussion of these concepts in the framework of
a low-energy effective theory is given in Ref.~\cite{Landsteiner:2016}. Clearly,
the same should apply to the chiral kinetic theory. This means that one should add the following topological
contribution to the electric four-current density \cite{Bardeen, Landsteiner:2013sja, Landsteiner:2016}:
\begin{equation}
\delta j^{\mu} =  \frac{e^3}{4\pi^2 \hbar^2 c} \epsilon^{\mu \nu \rho \lambda} A_{\nu}^5 F_{\rho \lambda},
\label{consistent-def-0}
\end{equation}
where $A_{\nu}^5=b_{\nu}+\tilde{A}^5_{\nu}$ is the axial vector potential, which is an observable quantity.
Indeed, in Weyl materials, $b_0$ and $\mathbf{b}$ correspond to the energy and momentum-space separations
between the Weyl nodes.
On the other hand, $\tilde{A}_{\nu}^5$ is expressed through the deformation tensor and
describes strain-induced axial (pseudoelectromagnetic) fields. [Note that there
is also a correction to the chiral current density, but it contains only pseudoelectromagnetic
fields \cite{Landsteiner:2016} and, thus, will not affect the plasmon properties, discussed later.]
In components, Eq.~(\ref{consistent-def-0}) takes the following form:
\begin{eqnarray}
\delta \rho &=&\frac{e^3}{2\pi^2 \hbar^2c^2}\,(\mathbf{A}^5\cdot\mathbf{B}),
\label{consistent-charge-density}
 \\
\delta \mathbf{j} &=&\frac{e^3}{2\pi^2 \hbar^2 c}\,A^5_0 \mathbf{B} - \frac{e^3}{2\pi^2 \hbar^2 c}\,(\mathbf{A}^5\times\mathbf{E}).
\label{consistent-current-density}
\end{eqnarray}
For $\mathbf{B}_{5}$ to be nonzero, the axial field $\tilde{\mathbf{A}}^5$
should depend on coordinates. We will assume, however, that such a dependence is weak and $\tilde{\mathbf{A}}^5$
in Eqs.~(\ref{consistent-charge-density}) and (\ref{consistent-current-density}) is negligible compared to the chiral shift $\mathbf{b}$.

As is easy to check, the consistent current ${J}^{\mu}
=(c\rho+c\delta \rho, \mathbf{j}+\delta \mathbf{j})$ is nonanomalous, $\partial_\mu {J}^{\mu}=0$, therefore, the electric charge
is locally conserved. Note that the consistent current plays an important
role even in the absence of strain-induced pseudoelectromagnetic fields. For example, in
the equilibrium state with $\mu_5=-eb_0$, the first term in Eq.~(\ref{consistent-current-density}) exactly cancels the
corresponding CME current in $\mathbf{j}$ as argued in Ref.~\cite{Landsteiner:2016}. This also agrees with
the analysis based on the band theory of solids \cite{Franz}. In addition, the second
term in $\delta \mathbf{j}$ correctly captures the anomalous Hall effect in Weyl materials \cite{Grushin-AHE}, that otherwise
would be missing in the chiral kinetic theory.

{\em Collective excitations.---}
By making use of the consistent current, we study the spectrum of collective excitations in the Weyl material
in a constant background field $\mathbf{B}_{0,\lambda}\equiv \mathbf{B}_0+\lambda \mathbf{B}_{0,5}$. For the
sake of simplicity, we assume that a static strain-induced pseudomagnetic field $\mathbf{B}_{0,5}$ is parallel to the
magnetic field $\mathbf{B}_0$. (We choose $\mathbf{B}_0$ to point in the $+z$ direction.) Note that, in principle,
collective modes could drive dynamical deformations of the Weyl material, which, in turn, induce oscillating
pseudoelectromagnetic fields $\mathbf{E}^{\prime}_5$ and $\mathbf{B}^{\prime}_5$. However, the corresponding
fields are extremely weak and can be safely neglected in the analysis of the plasmon modes.

Our analysis of the electromagnetic collective modes follows the standard approach of physical kinetics
\cite{Landau:t10}, albeit generalized to the case of chiral fermions with a nonzero Berry curvature. As usual, the solution
is sought in the form of plain waves, i.e.,
$\mathbf{E}^{\prime} = \mathbf{E} e^{-i\omega t+i\mathbf{k}\cdot\mathbf{r}}$ and
$\mathbf{B}^{\prime} = \mathbf{B} e^{-i\omega t+i\mathbf{k}\cdot\mathbf{r}}$, where
$\omega$ is the frequency and $\mathbf{k}$ is the wave vector. The matter effects
are captured by the polarization vector
\begin{equation}
P^{\prime m}=\chi^{ml}E^{\prime l}=  i J^{\prime m}/\omega,
\label{polarization-tensor}
\end{equation}
where $\chi^{ml}$ is the electric susceptibility tensor and $m,l=1,2,3$ are the spatial indices.
The dispersion relations
of the collective modes follow from the characteristic equation for the in-medium Maxwell's
equations \cite{Landau:t10}:
\begin{equation}
\mbox{det}\left[\left(n_0^2\omega^2- c^2k^2 \right)\delta^{lm} + c^2 k^l k^m + 4\pi\omega^2\chi^{lm}\right]=0,
\label{collective-B-tensor-dispersion-relation-general}
\end{equation}
where we included the background refractive index $n_0$. In the case of the Weyl semimetal TaAs, for example,
$n_0\approx6$ \cite{Buckeridge}. In this Letter, in order to simplify our analysis, we will neglect the dependence of
$n_0$ on the frequency. Also, we will discuss the properties of the collective modes only in the limit $\mathbf{k}=0$.
The general case with $\mathbf{k}\neq 0$ will be reported elsewhere. In order to determine
the electric susceptibility tensor, we use the consistent chiral kinetic theory, which includes the contribution to
the electric current due to the Bardeen-Zumino polynomial given by Eq.~(\ref{consistent-current-density}).
 The distribution function is taken in the form
$f_{\lambda}=f_{\lambda}^{\rm (eq)}+ f_{\lambda}^{\prime}$, where $f_{\lambda}^{\rm (eq)}$ is the equilibrium distribution function
(\ref{CKT-equilibrium-function}) and $f_{\lambda}^{\prime} = f_{\lambda}^{(1)} e^{-i\omega t}$ is a perturbation.
To leading linear order in oscillating fields, the solution to the kinetic equation (\ref{CKT-kinetic-equation}) reads
\begin{eqnarray}
f_{\lambda}^{(1)} &\simeq& -i\frac{ev_F}{p\omega}\frac{\partial f_{\lambda}^{\rm (eq)}}{\partial \epsilon_{\mathbf{p}}} \Bigg\{ (\mathbf{p}\cdot\mathbf{E})\left[1 +\frac{\lambda \hbar e (\mathbf{B}_{0,\lambda}\cdot\mathbf{p})}{2cp^3}\right]  \nonumber\\
&-&i\left(\mathbf{p}\cdot[\mathbf{B}_{0,\lambda}\times\mathbf{E}]\right) \frac{ev_F}{cp\omega}\Bigg\} .
\label{consistent-k=0-f-sol-2}
\end{eqnarray}
Similarly to the situation in a magnetized nonrelativistic plasma \cite{Landau:t10}, the leading-order perturbation
$f_{\lambda}^{(1)}$ is proportional to the magnitude of the oscillating electric field. By making use of this solution,
we derive the following result for the polarization vector:
\begin{eqnarray}
\mathbf{P}^{\prime}&=& \frac{a_0}{4\pi} \mathbf{E}^{\prime} +\frac{a_1}{4\pi} (\mathbf{b}\times \mathbf{E}^{\prime})
+\frac{a_2}{4\pi} (\mathbf{E}^{\prime}\times\hat{\mathbf{z}}),
\label{consistent-k=0-polarization-3}
\end{eqnarray}
where $\hat{\mathbf{z}}$ is the unit vector in the $+z$ direction and
\begin{eqnarray}
\label{consisten-k=0-a0}
a_0 &=&-\frac{ n_0^2 \Omega_e^2}{\omega^2}, \qquad a_1 = -i\frac{2e  n_0^2 \alpha v_F}{\pi c \omega \hbar}, \\
\label{consisten-k=0-a2}
a_2 &=&-i\frac{2e n_0^2 \alpha v_F^2}{3\pi\omega c} \sum_{\lambda=\pm}\left[\frac{B_{0,\lambda}\mu_{\lambda}}{\hbar^2\omega^2}
-\frac{B_{0,\lambda}}{4T}F\left(\frac{\mu_{\lambda}}{T}\right)\right].
\end{eqnarray}
Here we introduced the shorthand notations for the coupling constant $\alpha=e^2/(\hbar v_F n_0^2)$,
the Langmuir frequency
\begin{equation}
\Omega_e \equiv \sqrt{\frac{4\alpha}{3\pi\hbar^2}\left(\mu^2+\mu_5^2 +\frac{\pi^2 T^2}{3}\right)},
\end{equation}
and the following function of $\nu_\lambda \equiv \mu_\lambda/T$:
\begin{eqnarray}
\label{consistent-k=0-F1-def}
F\left(\nu_\lambda \right) \equiv -T\int\frac{dp}{p}\left\{ \frac{\partial f^{\rm (eq)}_{\lambda}}{\partial \epsilon_{\mathbf{p}}}
-\frac{\partial \bar{f}^{\rm (eq)}_{\lambda}}{\partial \epsilon_{\mathbf{p}}} \right\}.
\end{eqnarray}
Note that the high- and low-temperature asymptotes of this function are given by
$F\left(\nu_\lambda\right)\simeq 0.426\,\nu_\lambda $
for $\nu_\lambda\to 0$ and
$F\left(\nu_\lambda\right)\simeq 1/\nu_\lambda$ for $\nu_\lambda\to \infty$, respectively.

While the first term in Eq.~(\ref{consistent-k=0-polarization-3}) describes the high-frequency version of the Ohm's law,
the second term comes from the part of the topological current
in Eq.~(\ref{consistent-current-density}) responsible for the anomalous Hall effect \cite{Grushin-AHE}. The
last term in Eq.~(\ref{consistent-k=0-polarization-3}), which is proportional to the background magnetic and pseudomagnetic fields in view of Eq.~(\ref{consisten-k=0-a2}),
describes the usual Faraday rotation as well as its anomalous counterpart.

\begin{figure}[t]
\begin{center}
\includegraphics[width=0.9\columnwidth]{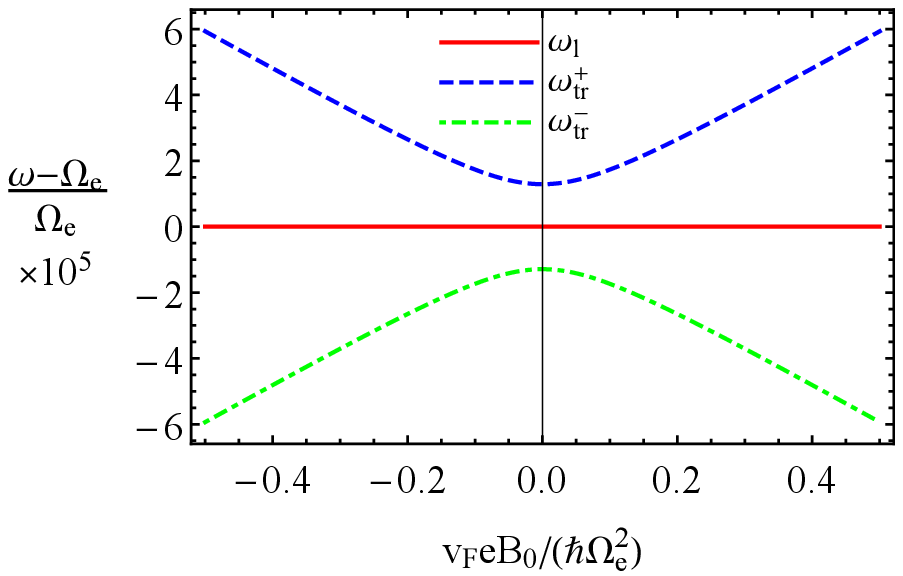}\vspace{2mm}
\includegraphics[width=0.9\columnwidth]{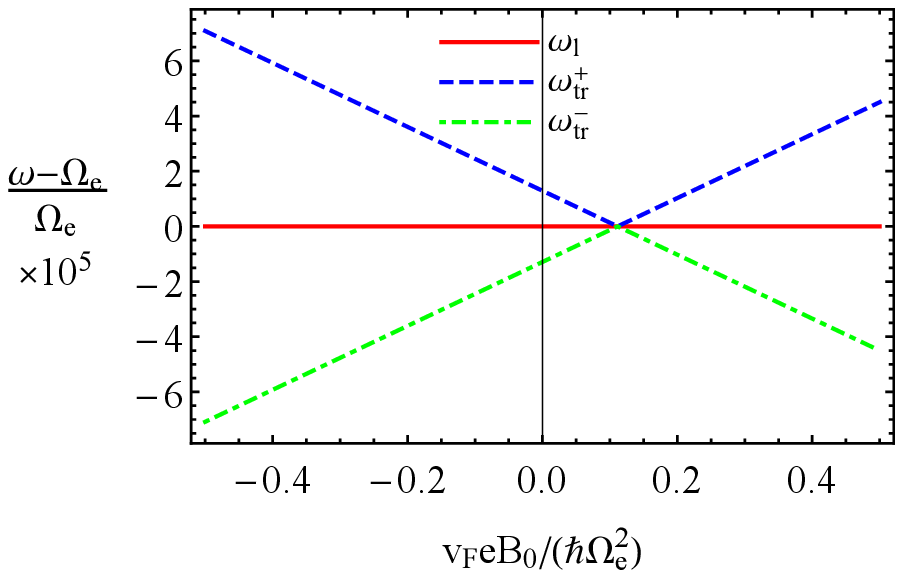}
\caption{(Color online) The dispersion relations of collective modes given by Eq.~(\ref{consistent-k=0-Sol}) at fixed
$b_{\perp}=0.2\,\hbar\Omega_e/e$ (upper panel) and $b_{\parallel}=0.2\,\hbar\Omega_e/e$ (lower panel). The electric chemical potential equals
$\mu=\hbar\Omega_e\sqrt{3\pi/(4\alpha)}$, $B_{0,5}=0$, and
the temperature is zero.}
\label{fig:consistent-k=0}
\end{center}
\end{figure}

By making use of Eqs.~(\ref{polarization-tensor}), (\ref{collective-B-tensor-dispersion-relation-general}), and (\ref{consistent-k=0-polarization-3}),
we obtain the spectral equation for the collective modes at $\mathbf{k}=0$
\begin{equation}
(n_0^2+a_0)\left\{ (n_0^2+a_0)^2+a_{1}^2 b^2_{\perp}+\left(a_2-a_1b_{\parallel}\right)^2\right\}=0,
\label{consistent-k=0-dispersion-tr}
\end{equation}
where we introduced the transverse $b_{\perp}=\sqrt{b_x^2+b_y^2}$ and longitudinal $b_{\parallel} =b_z$ components of the chiral
shift. Notice that the spectral equation is explicitly factorized. The corresponding approximate solutions are
\begin{eqnarray}
\omega_l=\Omega_e, \quad\quad
\omega_{\rm tr}^{\pm}=\Omega_e\,\sqrt{1 \pm \delta \Omega_e/ \Omega_e},
\label{consistent-k=0-Sol}
\end{eqnarray}
where
\begin{eqnarray}
\delta \Omega_e &=&\frac{2e \alpha v_F}{3\pi c \hbar^2} \Bigg\{9\hbar^2 b_{\perp}^2
+\Big[\frac{2v_F}{\Omega_e^2}(B_0\mu +B_{0,5}\mu_5) \nonumber\\
&-& 3\hbar b_{\parallel}
- \frac{v_F\hbar^2}{4T} \sum_{\lambda=\pm}B_{0,\lambda}F\left(\frac{\mu_{\lambda}}{T}\right)\Big]^2\Bigg\}^{1/2}.
\label{consistent-k=0-Sol-omega-pm}
\end{eqnarray}
In the absence of the chiral shift, the collective modes (\ref{consistent-k=0-Sol}) correspond to the longitudinal
($\mathbf{E}^{\prime}\parallel\hat{\mathbf{z}}$) and transverse ($\mathbf{E}^{\prime}\perp\hat{\mathbf{z}}$) waves.
Moreover, Eq.~(\ref{consistent-k=0-Sol}) means that the effects of dynamical electromagnetism transform, as argued in Ref.~\cite{Kharzeev}, the CMW into a
longitudinal plasmon, whose frequency coincides exactly with the Langmuir one at linear order in the (pseudo-)magnetic field.
It is interesting to point out that the combined effect of the pseudomagnetic field $\mathbf{B}_{0,5}$ and the chiral chemical
potential $\mu_5$ on the collective modes is similar to that of the magnetic field $\mathbf{B}_{0}$ and the electric chemical
potential $\mu$. The qualitative dependence of the plasma frequencies (\ref{consistent-k=0-Sol}) on the magnetic
field $B_0$ is presented graphically in Fig.~\ref{fig:consistent-k=0} at fixed values of $b_{\perp}$ and $b_{\parallel}$.

According to the upper panel in Fig.~\ref{fig:consistent-k=0}, the plasma frequencies of all three collective
modes are different when $b_{\perp}\neq0$. In this case, the smallest splitting occurs at $B_{0}=0$, where
$\omega_{\rm tr}^{+}-\omega_{\rm tr}^{-}\approx \delta \Omega_e =  2e\alpha v_F b_{\perp}/( \pi c\hbar)$.

The situation is quite different in the case when $b_{\perp}= 0$, but $b_{\parallel}\neq0$. This is demonstrated
in the lower panel of Fig.~\ref{fig:consistent-k=0}. Now, while the three plasmons have generically
different frequencies, one can make them degenerate by tuning the value of the magnetic field. The corresponding
value of the magnetic field $B_{0}^{\star}$, at which the frequency splitting vanishes, is given by
\begin{eqnarray}
B_{0}^{\star} &=& -\frac{4T\left[2v_F B_{0,5}\mu_5 -3\hbar\Omega_e^2 b_{\parallel} \right] }
{v_F\left[8T\mu -\hbar^2\Omega_e^2\sum_{\lambda=\pm}F\left(\frac{\mu_{\lambda}}{T}\right)\right]} \nonumber\\
&+& \frac{B_{0,5} \hbar^2\Omega_e^2\sum_{\lambda=\pm}\lambda F\left(\frac{\mu_{\lambda}}{T}\right) }
{8T\mu -\hbar^2\Omega_e^2\sum_{\lambda=\pm}F\left(\frac{\mu_{\lambda}}{T}\right)}.
\label{consistent-k=0-B0-crit}
\end{eqnarray}

{\em Chiral magnetic plasmons.}---
It is worth discussing the chiral features of the collective excitations in more detail. It appears that these modes,
including the longitudinal one, which describes the CMW with the effects of dynamical electromagnetism taken into account,
are chiral plasmons, or rather chiral magnetic plasmons, when a background magnetic field is present. Their chiral
nature is evident from the fact that they are accompanied by oscillations of not only the electric, but also the chiral current
density. The result for the oscillating part of the electric current density is clear from the polarization vector if one uses
Eqs.~(\ref{polarization-tensor}) and (\ref{consistent-k=0-polarization-3}). As for the oscillating part of the chiral current density, it is
given by the following expression:
\begin{eqnarray}
\label{collective-B-J5}
\mathbf{J}_5^{\prime} &=& \sin{(\omega t)}\mathbf{E}\,\frac{2\alpha  n_0^2 \mu\mu_5}{3\pi^2 \hbar^2\omega}
-\cos{(\omega t)} (\mathbf{E}\times\hat{\mathbf{z}}) \frac{e\alpha  n_0^2 v_F^2}{6\pi^2 c} \nonumber\\
&\times&\sum_{\lambda=\pm}\left[\frac{ \lambda B_{0,\lambda}\mu_{\lambda}}{\hbar^2\omega^2}
-\frac{\lambda B_{0,\lambda}}{4T}F\left(\frac{\mu_{\lambda}}{T}\right) \right],
\end{eqnarray}
which is obtained using Eqs.~(\ref{CKT-current}) and (\ref{consistent-k=0-f-sol-2}).
It is important to emphasize the topological origin of the first term in Eq.~(\ref{collective-B-J5}), which
does not depend on temperature. In essence, it comes from a dynamical version of the chiral electric
separation effect \cite{Huang:2013}. The second term in Eq.~(\ref{collective-B-J5}) is related to a generalized
Lorentz force.

We would like to note that the predicted frequencies and the splitting of plasmon frequencies
as functions of an applied strain and/or magnetic field can be easily tested in experiment. As in
the case of usual plasmons, this can be done by measuring the intensity and the phase shift of electromagnetic
waves transmitted through a thin film of a Weyl material. The frequencies of transverse modes
could be obtained from the peaks in the real part of optical conductivity, while the frequency of the
longitudinal mode can be extracted from the energy loss function (e.g., see Ref.~\cite{Pines}).

Depending on the choice of a Weyl material, the estimated frequencies of the chiral
magnetic plasmons could vary a lot. In Weyl semimetals such as NbP and TaAs, for example,
the averaged Fermi velocity is about $v_F\approx2 \times10^7~\mbox{cm/s}$ \cite{Lee:2015exa}. The
corresponding Langmuir frequency may vary in a rather wide range between $1~\mbox{THz}$
to $100~\mbox{THz}$ depending on the actual values of the Fermi energy and temperature.
The range of magnitude of the splitting between the transverse modes is more narrow, i.e.,
$\omega_{\rm tr}^+ -\omega_{\rm tr}^{-} \approx 0.3\, b_{\perp}[\mbox{\AA}^{-1}]~\mbox{THz}$,
where the value of the chiral shift parameter $b_{\perp}$ varies from about $4\times10^{-3}~\mbox{\AA}^{-1}$
(NbAs) to about $3\times10^{-2}~\mbox{\AA}^{-1}$ (TaAs) \cite{Lee:2015exa}.

{\em Conclusion.---} As we showed in this Letter, the consistent chiral kinetic theory in Weyl materials should necessarily
include the topological Chern--Simons contribution that ensures the local conservation of
the electric charge in electromagnetic and strain-induced pseudoelectromagnetic fields. Moreover, as we emphasized, such
a term plays an important role even in the absence of pseudoelectromagnetic fields. It allows one to correctly
describe the anomalous Hall effect in Weyl materials \cite{Grushin-AHE} and to reproduce the vanishing CME current in
an equilibrium state of chiral plasma \cite{Franz,Landsteiner:2016}. Furthermore,
the topological term also affects the spectra of collective modes.

As demonstrated here, the collective modes in Weyl materials are the chiral plasmons with interesting
properties. Such modes are associated with the oscillations of both electric and chiral current densities.
This is in contrast to the ordinary electromagnetic plasmons which are not connected with the oscillations of the
chiral current density. It is worth mentioning that for the longitudinal mode, which corresponds to the CMW, these oscillations are
of purely topological origin and are related to a dynamical version of the chiral electric separation effect.

While the plasma frequency of the longitudinal mode coincides with the Langmuir one, the frequencies
of the transverse modes generically split up. The frequency splitting depends on both magnetic
(pseudomagnetic) field and electric (chiral) chemical potential. As we showed, the qualitative features
of this dependence on the magnetic field can be used to develop a protocol for experimentally extracting
both the direction and magnitude of the chiral shift parameter in Weyl materials.

In this Letter, the study was restricted to the long-wavelength limit ($\mathbf{k}=0$)
of the chiral magnetic plasmons and used an expansion to the linear order in background
magnetic and pseudomagnetic fields. The generalization of this investigation to the case of nonzero wave vectors
($\mathbf{k}\neq 0$) and higher orders in magnetic and pseudomagnetic fields will be reported
elsewhere.

The work of E.V.G. was supported partially by the Program of Fundamental Research of the Physics and Astronomy Division of the NAS of Ukraine.
The work of V.A.M. and P.O.S. was supported by the Natural Sciences and Engineering Research Council of Canada.
The work of I.A.S. was supported by the U.S. National Science Foundation under Grant No.~PHY-1404232.

\end{document}